\documentclass[icps3]{svjour}
\usepackage{graphicx}

\begin{document}

\title{Spin-related transport in ultra small Si quantum dots}

\titlerunning{Spin-related transport in ultra small Si quantum dots}

\author{Leonid P. Rokhinson\thanks{e-mail: leonid@ee.princeton.edu}
\and L. J. Guo\thanks{\emph{Present address:} Department of
Electrical Engineering, University of Michigan, Ann Arbor, MI 48109
USA} \and D. C. Tsui \and S. Y. Chou}

\authorrunning{Leonid P. Rokhinson {\it et al.}}

\institute{Department of Electrical Engineering, Princeton
University, Princeton, NJ 08544 USA}

\maketitle

\begin{abstract}
We investigated electron transport through ultra small Si quantum
dots. We found that the $B$-dependence of energy levels is dominated
by the Zeeman shift, allowing us to measure the spin difference
between two successive ground states directly. In some dots the
number of electrons $N$ in the dot can be tuned starting from zero,
and the total spin of the dot can be mapped as a function of $N$ and
$B$.  For one of the dots we deduced that the dot becomes
spontaneously polarized at $N=6$ with a large spin change $\Delta
S=3/2$, demonstrating the essential features of spin blockade.
Surprisingly, for $N>20$, the transitions with $\Delta S>1/2$ do not
lead to the suppression of the corresponding peaks at low
temperatures.
\end{abstract}


\section{Introduction}

Recent advances in nanotechnology has made it possible to fabricate
quantum dots so small that Coulomb blockade can be observed up to
room temperature.  In such small dots, single--particle energies due
to the size quantization are comparable to the electrostatic charging
energy and interaction effects can modify the entire energy spectrum.
Important information about interactions in such quantum systems can
be obtained by studying the spin states.  However, it proved to be a
formidable task to measure the spin of a few-electron system
experimentally.
So far, the most successful mapping of spins
in a few-electron dot has been achieved indirectly, by comparing an
experimentally obtained addition spectrum to the theoretically
calculated energy spectrum for a particular
geometry\cite{ashoori93,tarucha96}.  The problem of measuring spin
directly in the most versatile and well studied vertical and lateral
quantum dots is that the magnetic field $B$ dependence of their
energy levels is dominated by orbital effects due to the weak
confinement of electrons by electrostatic gates.

In our small Si samples we are able to measure the spin of a
few-electron quantum dot directly. The dots uniquely combine the
flexibility to change the number of electrons $N$ starting from 0
with the strong electron confinement provided by the sharp Si/SiO$_2$
interface. At $B<13$ T, the $B$-dependence of energy levels is
dominated by the Zeeman shift and we are able to measure the
difference between the spin of the successive ground states $\Delta
S=S(N)-S(N-1)$ directly as electrons are added into the dot one by
one.  Also, we can follow the change of the spin configuration of the
dot with a fixed number of electrons as a function of $B$. For one of
the samples we can explain observed transition for the groundstates
with up to 5 electrons within a simple model of a few energy levels
which cross each other in magnetic filed as a result of Zeeman
splitting.  Some many-body effects, such as spontaneous dot
polarization and spin blockade\cite{weinman95} due to the spin change
$\Delta S>1/2$ has been observed.


\section{Samples}

The measurements were performed on small Si quantum dot fabricated
from a silicon-on-insulator wafer. The dot resides inside a narrow
bridge patterned from the top Si layer.
A 50 nm thick layer of thermal oxide is grown
around the bridge followed by a poly-Si gate. The fabrication steps
have been described previously in details\cite{leobandung95}. Spacing
between excited levels $\delta\sim 0.5-4$ meV, measured using
non-zero bias spectroscopy, is comparable to the charging energy
$U_c=e/C\approx 10$ meV and is consistent with the lithographical
size of the dot $l\approx\sqrt{\hbar/m^*\delta}\approx100-190$ \AA.

\section{Results and discussion}

Peak position as a function of $V_g$ is determined by the degeneracy
condition that the electrochemical potentials for the ground states
with $N-1$ and $N$ electrons in the dot are equal. It has long been
realized that for non-interacting electrons the field dependence of
the peak positions $V_g^p(B)$ can be directly mapped onto the
single-particle energy spectrum of the dot $E(N,B)$, provided that
the Fermi energy $E_F$ in the contacts is field independent.  For a
dot with a weak confining potential we expect energy levels to shift
by $~\hbar\omega_c$ or by the level spacing, whichever energy is
smaller, and the shift should strongly depend on the direction of the
magnetic field due to the dot anisotropy.

\begin{figure}
\includegraphics[width=\hsize]{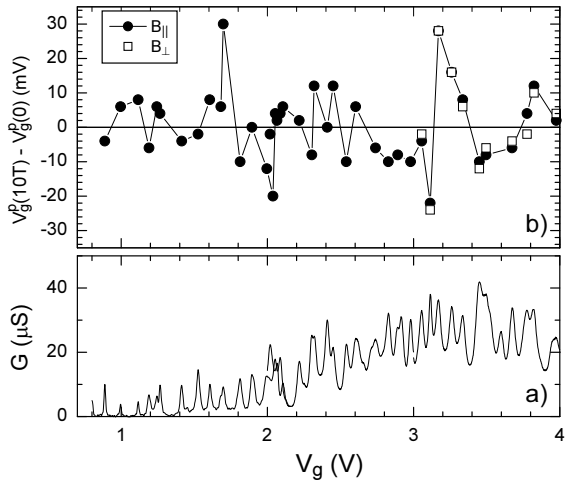}
\caption{a) Conductance as a function of the gate voltage for sample
E5-5B was measured at $T=1.5$ K and $B=0$. In b) a relative shift of
the peaks shown after an application of $B=10$ T.}
\label{shift}
\end{figure}

In Fig.~\ref{shift}a the conductance is plotted as a function of the
gate voltage for sample E5-5b.  The dot contains a few electrons (we
estimate $\approx 15$ electrons at $V_g=1$ V) and the number of
electrons can be changed by varying the gate voltage, each peak
corresponding to an electron added to or removed from the dot. In
Fig.~\ref{shift}b, the shift of each peak at $B=10$ T, relative to
its zero-field position, is plotted.  The average peak shift is much
less than $\hbar\omega_c=6$ meV ($\Delta V_g=52$ mV for this sample)
and is determined by the level spacing.  The most striking result is
that the shift is almost independent on the $B$ direction, which
suggests that spin effects dominate.

\begin{figure}
\includegraphics[width=\hsize]{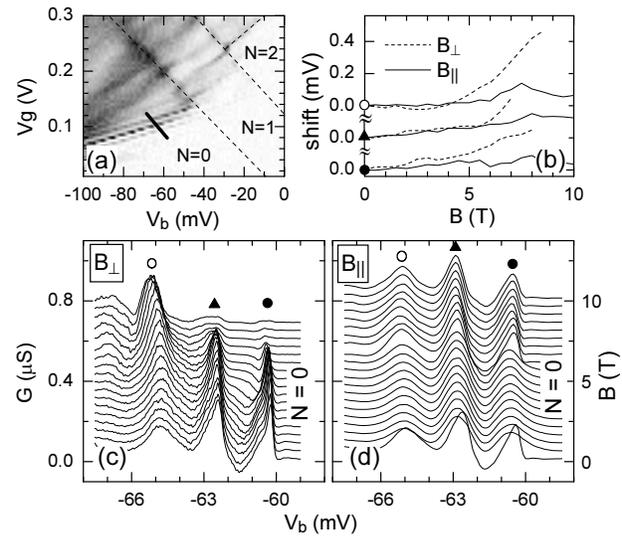}
\caption{a) High bias spectroscopy of a dot (E5-3b) is shown in a
gray scale plot of conductance.  White areas correspond to zero
conductance (Coulomb blockade for $N>0$) and dark lines are the
positions of the excited levels. In c) and d) conductance as a
function of bias, measured along the thick line in a), is plotted for
two configurations of the magnetic field: along the current direction
($B_{||}$) and perpendicular to the surface of the sample
($B_{\bot}$). The curves are offset by 0.08 $\mu$S/T.  The data in c)
and d) appears slightly different because they were taken at
different conditions: c) ac bias 30 $\mu$V and  $T=1.5$ K , d) ac
bias 100 $\mu$V and $T=1.5$ K, except the $0.5<B<6$ T curves which
were taken at 4 K. In b) shifts of the three peaks, marked by a dot,
a triangle and a circle in c) and d), are plotted as a function of
$B_{||}$ and $B_{\bot}$.  According to the capacitance model the
conversion factor for the shift into energy $\beta=e\Delta V_b/\Delta
E=2$. However, the slope of the lowest-$V_g$ peaks in the $V_g-V_b$
plane is increased by a factor of 4, compared to the slopes for the
peaks with $N>1$, suggesting that it requires an extra bias to force
the first electron into the dot. Thus, we estimate $\beta\sim8$.}
\label{spectr}
\end{figure}

It is possible to separate orbital and spin effects if contacts are
spin polarized, as shown in Fig.~\ref{spectr}. In the plot an
excitation spectrum of another dot, E5-3b, is shown for the entrance
of the first electron. At low gate voltage $V_g<0.1$ V the contact is
spin polarized for $B>1$ T and the Zeeman shift of the first energy
level is fully compensated by the Zeeman shift of the Fermi energy in
the contacts. Thus, the resulting shift is solely due to orbital
effects (assuming that $g$-factors in the dot and in the contacts are
the same).  The net shift of the levels in a parallel field (the
smallest cross section) is negligible, while there is a small but
measurable parabolic dependence on $B_{\bot}$ (the largest cross
section).  The net shift at 10 T is $<0.4$ meV and Zeeman energy
should dominate the $B$ dependence ($\frac{1}{2}g^*\mu_B B=0.6$ meV
at $B=10$ T).
It is interesting to notice, that although orbital effects are not
effecting the position of the energy levels appreciably, they can
considerably change transmission coefficients. For example, at high
perpendicular field the peaks, marked with the dot and triangle in
Fig.~\ref{spectr}, become suppressed, while the third peak is
enhanced, compared to the $B=0$ value.

For a larger number of electrons $N>3$ the contacts are
spin-degenerate up to the highest experimental field of 13 T and the
shift of energy levels mainly reflects the Zeeman shift of the
tunneling electron. Peaks shift with a slope $\frac{1}{2}g^*\mu_B$ as
a function of $B$ and the slope changes sign every time the two
levels with different spin cross each other.  Evolution of the peaks
for up to 5 electrons in the dot can be understood within a model of
non-interacting electrons and singlet-triplet and triplet-polarized
transitions are identified as a function of $B$. Peak 6 has a large
slope $\approx \frac{3}{2}g^*\mu_B$ at low $B$ and is strongly
suppressed at low $T$. We conclude that the $N=6$ ground state is
spontaneously polarized at $B=0$ and the peak suppression is due to
the long predicted spin blockade\cite{weinman95}.  For larger numbers
of electrons, $N>20$, we observed transitions with $\Delta S=3/2$
with no apparent suppression of the peaks at low $T$.  The spin
scattering mechanism involved is not currently understood. These data
will be published elsewhere\cite{rokhinson00b}.

The work was supported by ARO, ONR and DARPA.


\end{document}